# Recommendations for extending the GFF3 specification for improved interoperability of genomic data

## Authors

Surya Saha (1), Scott Cain (2), Ethalinda K. S. Cannon (3), Nathan Dunn (4), Andrew Farmer (5), Zhi-Liang Hu (6), Gareth Maslen (7), Sierra Moxon (8), Christopher J Mungall (8), Rex Nelson (3), Monica F. Poelchau (9)

((1) Boyce Thompson Institute, Ithaca, NY, (2) Ontario Institute for Cancer Research, Toronto, Canada, (3) USDA, Agricultural Research Service, Corn, Insects, and Crop Genetics Research Unit, (4) Truveta, Seattle, WA, (5) National Center for Genome Resources, (6) Iowa State University, Ames, IA, (7) European Molecular Biology Laboratory, European Bioinformatics Institute, Wellcome Genome Campus, Hinxton, UK, (8) Lawrence Berkeley National Lab, Berkeley, California, USA, (9) USDA, Agricultural Research Service, National Agricultural Library)

## Abstract

The GFF3 format is a common, flexible tab-delimited format representing the structure and function of genes or other mapped features (https://github.com/The-Sequence-Ontology/Specifications/blob/master/gff3.md). However, with increasing re-use of annotation data, this flexibility has become an obstacle for standardized downstream processing. Common software packages that export annotations in GFF3 format model the same data and metadata in different notations, which puts the burden on end-users to interpret the data model.

The AgBioData consortium is a group of genomics, genetics and breeding databases and partners working towards shared practices and standards. Providing concrete guidelines for generating GFF3, and creating a standard representation of the most common biological data types would provide a major increase in efficiency for AgBioData databases and the genomics research community that use the GFF3 format in their daily operations.

The AgBioData GFF3 working group has developed recommendations to solve common problems in the GFF3 format. We suggest improvements for each of the GFF3 fields, as well as the special cases of modeling functional annotations, and standard protein-coding genes. We welcome further discussion of these recommendations. We request the genomics and bioinformatics community to utilize the github repository (https://github.com/NAL-i5K/AgBioData_GFF3_recommendation) to provide feedback via issues or pull requests.

# Introduction

The GFF3 format is a commonly used tab-delimited format representing the structure and function of genes or other mapped features (https://github.com/The-Sequence-Ontology/Specifications/blob/master/gff3.md). The format's flexibility allows scientists to easily manipulate GFF3 files, and it helps accurately represent the complex biological information being captured. However, with increasing re-use of annotation data, in particular from different sources (software output from custom datasets, and/or reference datasets provided by databases), this flexibility has become an obstacle for downstream processing. Common software packages that export annotations in GFF3 format model the same data and metadata in different notations, which puts the burden on end-users to understand possibly undocumented assumptions about the data model, then to convert the data for downstream applications. For example, the CDS phase field is commonly misinterpreted by both dataset generators and consumers, which can lead to vastly different and erroneous amino acid sequences derived from the same GFF3 file.

The AgBioData consortium (https://www.agbiodata.org) is a group of genomics, genetics and breeding databases and partners working towards shared practices and standards[1]. Almost every AgBioData database uses the GFF3 format in some capacity, either for content ingest (into the database or associated tools, such as JBrowse[2]), analysis, distribution, or all of the above. AgBioData members report that much of their data wrangling time is spent reformatting and correcting GFF3 files that model the same data types in different ways. Providing concrete guidelines for generating GFF3, and creating a standard representation of the most common biological data types in GFF3 that would be compatible with the most commonly used tools, would provide a major increase in efficiency for all AgBioData databases.

The AgBioData GFF3 working group has developed new recommendations to solve common problems in the GFF3 format. We have referred to and in some cases adopted guidelines developed by the Alliance of Genome Resources (https://docs.google.com/document/d/1yjQ7lozyETeoGkPfSMTAT8IN3ZIAuy5YkbsBdjGeLww/edit), and NCBI (https://www.ncbi.nlm.nih.gov/datasets/docs/v1/reference-docs/file-formats/about-ncbi-gff3/; https://www.ncbi.nlm.nih.gov/sites/genbank/genomes_gff/). Below, we suggest improvements for each of the GFF3 fields, as well as the special cases of modeling functional annotations, and standard protein-coding genes. We welcome debate and discussion of these recommendations from the larger community - these recommendations will only be helpful if they are refined and then adopted by many. Our goal is to clarify the GFF3 specification and limit ambiguity for AgBioData and other databases and resources.

Table 1: Summary of recommendations

| Column | Change level | Attributes | Change level |
|---|---|---|---|
| Seqid (column 1) | Recommendation | ID | Recommendation |
| Source (column 2) | No change | Name | Recommendation |
| Type (column 3) | No change | Alias | Recommendation |
| Start, end (column 4, 5) | No change | Dbxref | Recommendation |
| Score (column 6) | Moderate | Derives_from | Recommendation |
| Strand (column 7) | No change | Note | No change |
| Phase (column 8) | Recommendation | Ontology_term | Recommendation |
| Modeling protein-coding genes | Recommendation | Target, Gap | Recommendation |
| | | Functional annotations | Major change |

## Specific recommendations

We recommend that developers and databases follow the Sequence Ontology GFF3 specifications (https://github.com/The-Sequence-Ontology/Specifications/blob/master/gff3.md) with emphases and additions below. Each field contains information in the following categories:
- Change level: The level of change relative to the SO specification. Values are 'No change', 'Recommendation only','minor', 'moderate', 'major'
- Summary: A summary of the GFF3 working group's findings.
- Proposed changes to specification: A list of the proposed changes to the SO specification.
- Rationale: The rationale behind these changes.
- Best Practices: Recommended best practices for this field.
- Validation: How software would validate whether the field is used correctly.
- Example: An example implementation of the field. Examples are listed in the Appendix.

1. ## Seqid (column 1)
   - **Change Level.** Recommendation only

- **Summary.** Optionally, provide an Alias table to specify alternate identifiers/aliases for the seqid.
- **Proposed Changes to Specifications**. Institute a new pragma for alias table link.
- **Rationale.** Sequences often have aliases (multiple identifiers, human-readable names that are not globally unique), and users prefer human-readable display names when viewing sequences in browsers.
- **Recommendation**. Optionally provide a machine- and human- readable 'alias' table to specify identifiers and their aliases, which is provided by GenBank, and requires INSDC submission of the genome.
- **Validation.**
    - A pragma line in the GFF3 header, beginning with ##alias-table, should provide a resolvable URL to the GenBank Alias table. Validator should only verify that the link is active.
    - If the GenBank Alias table is not available, then a separate alias table can be provided. A pragma line, ##alias-table [columns] indicates where the table begins, and the definitions of the columns provided. All identifiers in column 1 of the GFF3 file must be uniquely present in column 1 of the alias table. The columns in the alias table should be tab delimited. There will be no validation for an alternate alias table.
- **Example** in Supplementary data

# 2. source (column 2)
- **Change level:** no change
- **Summary**: There are no major changes from the previous SO specification. We recommend that the source field is used to define the source of the sequence feature concisely. Source is used to extend the feature ontology by adding a qualifier to the type field.
- **Proposed changes to specification:** none
- **Rationale**: The values used for this field vary widely as it's a free text field which can lead to parsing and interpretation issues for downstream software and data loading.
- **Best Practices**:
    - We recommend that programs generating and consuming GFF3 follow the constraints outlined below and account for the fact that the feature can be a result of multiple tools in a pipeline.
    - Optionally, a pragma can specify the source. We recommend following the VCF specification for Info/ID: https://samtools.github.io/hts-specs/VCFv4.3.pdf. We discourage verbose use of this pragma.
- **Validation:**
    - It is not necessary to specify a source. If there is no source, put a "." (a period) in this field.
    - Note that only spaces are allowed to represent whitespace. In general, follow the formatting requirements of the specification. From the GFF3 specification:

"Literal use of tab, newline, carriage return, the percent (%) sign, and control characters must be encoded using RFC 3986 Percent-Encoding; no other characters may be encoded."
- If there are multiple sources, use a literal comma to separate them (NOT %2C). Source names should not include literal commas.
- Specify the tool, method or pipeline used to generate this annotation or the database it was acquired from. Mention the version number if available.
- The feature should be a well defined output of the tool or database specified. If there is any ambiguity or post-processing, it should be clearly explained in an optional pragma stanza.
- The pragma will not be validated.
- **Example** in Supplementary data

## 3. type (column 3)

- **Change level**: no change
- **Summary:** We endorse the Alliance recommendations for the 'type' field when modeling hierarchical gene features. This aligns with the SO specification that expects this to be "either a term from the Sequence Ontology or an SO accession number".
- **Proposed changes to specification**: none
- **Rationale:** Software interpreting the type column can run into difficulties with complex cases. Software is easier to develop and maintain if we can make some simplifying assumptions about how genes are typically modeled. Using simple terms will additionally improve human readability and interpretation.
- **Best practice:** Top-level feature types can include 'gene' and 'pseudogene'. Optionally include a so_term_name attribute in column 9 to specify the child (type) of gene - e.g. protein_coding_gene, ncRNA_gene, miRNA_gene and snoRNA_gene (http://purl.obolibrary.org/obo/SO_0000704). Transcript features should include the appropriate SO term in column 3 (e.g. mRNA, snoRNA, etc).
- **Validation:**
    - Must be a valid SO term or SO accession number
    - All child rows should use a type within the hierarchy of the parent
    - A list of the SO terms and the hierarchy in OBO format is fetched from http://song.cvs.sourceforge.net/viewvc/*checkout*/song/ontology/sofa.obo by default
- **Example** in Supplementary data

## 4. start, end (column 4,5)

- **Change level:** no change
- **Summary**: No changes from the SO specification.
- **Proposed changes to specification:** none
- **Rationale**: Consistent use of *start* and *end* coordinates are essential.

- **Best practice**: We recommend that programs generating and consuming GFF3 be aware of the existence of circular chromosomes, which will require alternate interpretation of the end coordinates.
- **Validation**:
    - *start* and *end* are 1-based coordinates.
    - *start* must always be less than or equal to *end*.
    - A feature with no length, for example, an insertion site, is indicated by *start = end*. The insertion site is to the right of the position. There is no recommendation for representing an insertion at the beginning, that is, before the first base as 0 is an invalid coordinate.
    - A feature that is one base in length, e.g., a SNP, is also indicated by start = end. Distinguishing between one-base and zero-length features will have to rely on other fields, such as the type field.
    - In the absence of the *'Is_circular=true'* attribute in column 9, *end* indicates the terminal coordinate of the feature.
    - If *'Is_circular=true'* appears in column 9, *start* gives the beginning coordinate of the feature and *end* is start + (feature length - 1). This means the value for *end* may be larger than the chromosome size
- **Example** in Supplementary data

5. **score (column 6)**
    a. **Change level:** moderate
    b. **Summary**: There is no clear guidance on how to interpret the score column. Therefore, define how the score was calculated in a pragma.
    c. **Proposed changes to specification:** Define score calculation via pragma.
    d. **Rationale**: There is currently no standard for providing metadata or context for the score column, rendering the score essentially meaningless.
    e. **Best practice**: Optionally, define how the score was calculated in a pragma. May have multiple parts, such as score name, program, version, range, and whether quality increases or decreases or is constant with increasing values. Use EDAM ontology (https://edamontology.org) where possible. The score itself must be a floating point number. This recommendation considers the score column only when representing gene models.
    f. **Validation**:
        i. A period indicates no score.
        ii. If any record has a value in the score column:
            1. It must be a floating point number
            2. Optionally, there is a ##score pragma in the following format:
                a. ##Score name="[name/calculated-by]";min=[min-value]; max=[max-val];best=[lower/higher]
    g. **Example** in Supplementary data

# 6. Strand (column 7)

- **Change level:** No change.
- **Summary:** No change from the original specification.
- **Rationale**: NA
- **Best practices:** Follow the original specification.
- **Validation**: Values should include '+', '-', '.'. '?' can be used when the strand is relevant but unknown.
- **Example**: NA

# 7. phase (column 8)

- **Change level:** recommendation only
- **Summary**: Programs generating and consuming gff3 should pay close attention to the phase field and validate it, as phase is often incorrect.
- **Proposed changes to specification:** none
- **Rationale**: We have identified three main problems with the phase field. 1) Phase is often ignored or misinterpreted, both by programs generating gff3 and programs that consume it. When recorded manually, phase is often incorrect. This is problematic for programs that calculate the CDS and protein sequence using the combination of CDS coordinates and phase. Other methods that are frequently used to calculate the CDS and protein sequence (e.g. longest ORF, identifying start and stop codons) make critical assumptions that can also generate incorrect sequence, in particular for fragmented genomes where gene models may not have start and/or stop codons. 2) Even if the phase is correct, a translation table is required to correctly calculate the protein sequence, and there may be multiple translation tables needed for a given gff3, for example when both nuclear and organellar sequence is represented. 3) Even when the phase and translation tables are correct, the correct sequence may not be inferred due to post-translational modifications (e.g. selenocysteines) or problematic reference genome assemblies. NCBI represents these edge cases via the 'transl_except' attribute.
- **Best practices**:
    - We recommend that programs generating and consuming gff3 pay close attention to this value and validate it; however, validation may still fail in complex cases. Phase may be an example where the GFF3 format has reached its limit. In cases where the correct sequence may not be inferred due to post-translational modifications (e.g. selenocysteines) or problematic reference genome assemblies, use the 'transl_except' convention developed by NCBI on the CDS feature (transl_except=(pos:<base_range>%2Caa:<amino_acid>); https://www.ncbi.nlm.nih.gov/genbank/genomes_gff/)
    - Optionally provide a phase pragma, if any sequences in the GFF3 file do not use the standard genetic code (id = 1, see https://www.ncbi.nlm.nih.gov/IEB/ToolBox/C_DOC/lxr/source/data/gc.prt#0107 ). The pragma should provide the translation table id, and the reference sequences

in the GFF3 that will use that translation table id, e.g. ##phase <RefSeq ID> <translation table ID>;
- **Validation**: For a description of what phase means in the context of a single CDS line, see the 'Column 8: "phase' section of the current [GFF3 specification.](#) Optionally provide the translation table id, and the reference sequences in the GFF3 that will use that translation table id, in a phase pragma. The validator will use the translation tables in [https://www.ncbi.nlm.nih.gov/IEB/ToolBox/C_DOC/lxr/source/data/gc.prt](https://www.ncbi.nlm.nih.gov/IEB/ToolBox/C_DOC/lxr/source/data/gc.prt). If no phase pragma is given, or if not all reference sequences are specified in the phase pragma, the validator will use the Standard genetic code (id 1). Protein and CDS fasta are optional but highly recommended. Validator will generate CDS/protein sequences based on the phase specified in the gff3 file.
    - If no CDS/protein fasta is available:
        - Check for internal stops in the protein sequence - validation fails if stops are present
    - If CDS and protein fasta are available:
        - Compare given sequence to sequence generated from gff3 - validation fails if sequences are not identical
- **Example** in Supplementary data for the following use cases
    - Pragma specifying translation table: specify exemptions to standard code only
    - Example of incorrect vs. correct phase

# 8. attributes (column 9): ID
- **Change level:** recommendation only
- **Summary.** The ID attribute's role is to specify relationships between parent and child features within the GFF3. However, it is often - but not always - also used to specify a globally unique, persistent identifier. This second interpretation causes many problems with downstream software and validators. We recommend NOT using the ID attribute to specify the globally unique, persistent identifier, but instead using a separate attribute, such as Dbxref or gene_id.
- **Proposed changes to specification:** Recommend additional attributes (reserved or non-reserved) to specify the globally unique, persistent identifier
- **Rationale.** The GFF3 specification requires an ID attribute to define parent-child (part of) relationships for hierarchically modeled features (see specification for a detailed definition). In practice, the ID attribute is often also used to stand in for a globally unique, persistent identifier (in particular for genes, transcripts and proteins; see this article for definitions of and best practices on persistent identifiers: [https://journals.plos.org/plosbiology/article?id=10.1371/journal.pbio.2001414](https://journals.plos.org/plosbiology/article?id=10.1371/journal.pbio.2001414)). This second interpretation becomes problematic when users or software assume it is true in all cases, as they expect additional meaning beyond a generic, unique string. While a dual-use ID attribute may seem convenient, it is not always clear which level of a gene feature may have the 'true' persistent identifier - the gene, transcript, protein, even exon? In addition, it is not always possible to accurately model parent-child relationships if it is also a globally unique, persistent identifier.
- **Best practices.** Use the ID attribute as originally intended by the GFF3 specification, and do not assume that it contains a globally unique persistent identifier. For these, use an additional attribute. The attribute may depend on the use case - for example, if the persistent identifier is maintained by another database, Dbxref may be used (however, it may be confusing if multiple Dbxref identifiers are specified). The Alliance for Genome Resources uses the unreserved attributes gene_id, transcript_id, and protein_id, where the values of these are curies (https://en.wikipedia.org/wiki/CURIE). NCBI's RefSeq uses the unreserved attributes gene, transcript_id, and protein_id. We recommend using either of these conventions, with an additional 'feature_id' attribute for features that are neither genes, transcripts, or proteins. While this recommendation may require software and databases to adjust, this is simpler than forging a way for the ID attribute to meet all downstream tools' and users' needs. Note the still unresolved yet related problem of the corresponding FASTA definition line - there are no guidelines in terms of correspondence between the FASTA defline and identifying information in the GFF3 file.
- **Validation.** ID attribute: only validate whether IDs for each feature are unique within the scope of the GFF file, with the exception of discontinuous features. Persistent identifiers specified via the Dbxref attribute can be validated according to Dbxref rules.
- **Example** in Supplementary data

# 9. attributes (col 9): Name

- **Change level:** recommendation only
- **Summary:** A designation for the given feature used for display.
- **Proposed changes to specification**: None
- **Rationale:** Naming standards exist and should be followed when possible.
- **Best practices:**
    - Various taxonomic communities have nomenclature standards that aim to provide consistent naming across genes. Please identify and follow these standards for your community.
    - Name should not be mistaken for a unique identifier, or dbxref
    - Note that different, non-reserved attributes are sometimes used instead of Name. For example, NCBI uses product for the protein product name, gene_desc for the full gene name, and symbol as the gene abbreviation.
- **Validation:** refer to different community standards. No automated validation currently possible.
- **Example** in Supplementary data

# 10. attributes (col 9): Alias

- **Change level:** Recommendation only

- **Summary:** There are no major changes from the previous SO specification. The primary function of Alias appears to be for human consumption, display, indexing, and tracking synonyms or prior names. If there are additional functions that you know the value should be used for, we recommend that you choose a different attribute that is more specific, and could be consumed programmatically. The validator won't check Alias values.
- **Rationale:** The Alias field is used for alternate names or identifiers, and there are no real constraints on what these may be. Some important cases are as follows. If a gene is merged, or if the type of a genomic feature is changed, the name of the original feature may need to change. Conversely, if a gene is split, we should retain a reference to its original name. If two papers cite the same gene using a different name, we should be able to search for either one even if only one is an official one.
- **Proposed changes to specification**: None
- **Best practices:** Use Alias for human consumption of alternate or historical names and identifiers (e.g., gene merge), but do not assume that this field will be consumed programmatically. Alias should not be a replacement for Dbxref, and valid CURIE in Dbxref should be housed within Dbxref and not in Alias. Commas, tabs, and pipes should be avoided in alias names. It is possible to have multiple Alias values. It is recommended that these be separated via a comma.
- **Validation:** Any set of symbols would be appropriate and does not need to be unique. Alias values can not include a semicolon.
- **Example** in Supplementary data

## 11. attributes (col 9): Dbxref

- **Change level:** Recommendation only
- **Summary:** 1) Use the Dbxref field to cross-reference the *same* entity at a database - the field is sometimes mis-interpreted to reference related information, but not the same entity. 2) The Dbxref should result in a resolvable URL.
- **Proposed changes to specification:** Recommend use of the field for global entity references and crosslinks between databases.
- **Rationale:** The database cross-reference (dbxref) links a particular feature in the GFF record to a specific external database record by identity, source, association, or ontology links. To date, there are four established lists with hundreds of registered databases that offer external links by dbxref. The GO consortium list is maintained on GitHub with defined schema and format validation tools. It currently lists 266 databases (http://amigo.geneontology.org/xrefs). The UniProt Knowledgebase cross-reference list contains 183 databases (https://www.uniprot.org/docs/dbxref) flagged to represent 18 categories of databases (https://www.uniprot.org/database/) for better utility. The NCBI-GenBank db_xrefs list was developed in 1997 (https://doi.org/10.1038/ng0497-339), however it has only 129 databases listed to date (https://www.ncbi.nlm.nih.gov/genbank/collab/db_xref/). In addition, identifiers.org (https://identifiers.org/) provides a free service for looking up and referencing a data ID to one of the 714 pre-curated life science database locations using Compact Identifiers syntax. For example, the uniform resource identifier (URI) http://identifiers.org/pdb/2gc4 can be instantly forwarded to https://www.rcsb.org/structure/2gc4. The syntax of Compact Identifiers includes three parts: a provider code, a namespace prefix, and an accession. The provider code and namespace prefix are manually curated and stable, and can be easily looked up at its web site. Note that the databases represented by these four resources may overlap.
- **Best practices:**
    - The dbxref must refer to the same entity (not related information) in an external database. The format of a dbxref record may take the form "dbxref=database + identifier", where "database" is an abbreviated database name registered on a known dbxref list (above), affixed without space with a specific path leading to the database agent that accepts the "identifier" for information retrieval. The "database:identifier" construct is called unique resource identifier (URI). The URI must be specific to a record. Both dbxref record and the resolved URL should be a continuous string compliant to RFC-3986 [https://datatracker.ietf.org/doc/rfc3986/]. The composed URL must be unique and exist. Multiple dbxref items may be allowed for one GFF record. We recommend using identifiers.org for identifier resolution.
- **Pragma**: format: dbxref=URI; where URI is concatenated (without space) with the follow components:
    - Protocol://domain name/ + [namespace] + identifier
    - Name space is the path to the identifier handler;
    - Identifier: unique accession of an entity

- **Validation:** There may be a challenge when writing the validators, given that there are multiple independently maintained lists, and there may be lags in information updates. The validator could check whether a HTTP response status code 200 is returned based on the url built from the dbxref registry in the directive. Semicolons (";") are not permitted in the URL as it's used as the delimiter between column 9 "name=value" pairs..
- **Example**: See section "18. Progamas" for details.

## 12.     attributes (col 9): Derives_from
- **Change level:** minor
- **Summary:** The most common use for the Derives_from attribute is to describe the relationship between CDS and polypeptide features. However, 1) not all software recognizes this relationship, and 2) we do not recommend modeling polypeptide features in GFF3 (see recommendations for 'Modeling hierarchical relationships of a protein-coding gene'). Avoid modeling polypeptide features in general to prevent downstream interpretation problems of Derives_from.
- **Proposed changes to specification:** None.

- **Rationale:** The Derives_from attribute (http://purl.obolibrary.org/obo/RO_0001000) is used in situations where the relationship between features is temporal, and therefore the part_of relationship implied by the 'Parent' attribute is not appropriate (e.g. polypeptides are derived from CDS features, or in the case of polycistronic genes). In practice most programs that consume or create GFF3 do not check whether implied part_of relationships are actually valid per the Sequence Ontology (a notable exception is Genometools (http://genometools.org/cgi-bin/gff3validator.cgi)).
- **Best practices:** To avoid breaking software that consumes GFF3, we recommend not specifying a polypeptide feature if you're modeling a typical protein-coding gene based on genomic coordinates (see also recommendations for 'Modeling hierarchical relationships of a protein-coding gene' below).
- **Validation:** This needs further analysis and discussion.
- **Example** in Supplementary data

## 13. attributes (col 9): Note
- **Change level:** No change
- **Summary:** No changes relative to the original definition.
- **Proposed changes to specification**: None
- **Rationale:** Sometimes the Note attribute can contain irrelevant information.
- **Best practices:** Use primarily for notes relevant for public consumption (e.g., important errata for downstream users). Use as a last resort and at your own risk. In some cases a custom tag may be more appropriate. A common use-case would be for curation notes relevant to external users.
- **Validation:** May not include a semicolon. May be repeated.
- **Example** in Supplementary data

## 14. attributes (col 9): Ontology_term
- **Change level:** Recommendation only.
- **Summary:** Avoid.
- **Proposed changes to specification**: None
- **Rationale:** See section about functional annotation and metadata below.
- **Best practices:** Avoid. In general, do not use gff3 for functional annotation if possible. Use GAF, GPAD or other instead. If you do have to use it, use a CURIE backed ontology term (e.g. GO:0000077). In general, supplying an ontology term without underlying evidence, reference, or other context will be under-defined functional annotation and will be incomplete for downstream users and of little utility.
- **Validation:** Should be a CURIE that resolves into a published ontology term. The validator should issue a warning that Ontology_term should be avoided.
- **Example** in Supplementary data

## 15. attributes (col 9): Target, Gap

- **Change level:** Recommendation only.
- **Summary:** Target is an attribute intended to encode a parseable relationship between a region on the sequence given in column 1 and a region on another sequence, possibly (but not necessarily) another sequence referenced in column 1 of another record in the same gff file. Gap is an associated attribute encoding the alignment (i.e. gapping structure) needed to put the elements of the two sequences into homologous correspondence.
- **Proposed changes to specification**: None
- **Rationale:** The GFF3 specification describes the purpose of these two attributes as being for the representation of sequence alignments. Since the time that specification was written, numerous alternative formats for storing sequence alignments have been developed in response to the proliferation of sequence data brought about by the advent of next generation sequencing technologies. Therefore, although there may be cases in which the use of such attributes to represent alignments in the context of GFF files is convenient for a specific purpose (e.g. a gene prediction program seeking to represent the evidence behind its models), in general we advocate the use of alternative file formats such as BAM or PAF for representing alignments between sequences.
- There are additional contexts in which the representation of homology between regions on sequences is desirable without the fine grained structure of base pair correspondence. For example, syntenic relationships (SO:0005858) between regions of chromosomes that are derived from collinear blocks of homologous genes may be computed without having the full details of the alignment of the genomic regions. In this case, since the relationship is likely going to be an inter-genomic comparison (e.g. between species), it is important to be able to represent the information as a single record bearing the information about the paired genomic regions. Target can be used in such cases to represent the relationship, following the space-delimited structure suggested in the GFF3 specification "target_id start end [strand]". We note that although the GFF3 specification indicates that spaces in the target_id must be hex-encoded, we would recommend that the target_id represent a sequence identifier such as would be found in Column 1 of the same (or another) GFF3 file containing annotations of the target sequence.
- **Best practices:** We advocate the use of alternative file formats such as BAM or PAF for representing alignments between sequences.
- **Validation:** The components of the encoded attribute must be single-space delimited and must consist of not less than 3 and not more than 4 fields. Field 1 must conform to the same syntax and semantics as specified for Column 1 (seqid). Fields 2 and 3 must conform to the syntax and semantics for Columns 4 and 5 (start and stop) respectively. Field 4 is optional and conforms to the syntax and semantics of Column 7 (strand), though a missing value will be indicated by absence of the field rather than using ".".
- **Example** in Supplementary data

# 16. Attributes (col 9): complex metadata (e.g. functional annotations)

- **Change level:** Major change
- **Summary:** In some instances we may need to model complex sets of metadata within the GFF3.
- **Proposed changes to specification:** Provide a format for including richer metadata.
- **Rationale:** In general, modeling functional annotations in GFF3 should be avoided if other mechanisms ([GPAD](), [GPI](), [all spec formats and versions]() ) are available. However, in some instances other formats are not sufficient or readily available for tooling. In those instances, the ability to include information such as functional annotations that track annotation provenance, gene products, or properly annotated GO evidence may be necessary.
- **Best practices:**
    - We discourage the use of GO terms, or any functional annotation that requires an evidence code, without supplying the evidence code.
        - GO term or functional annotations should never be incorporated into gff3 within the Dbxref or Ontology_term fields. Another file format should be used, e.g. GAF or GPAD, if possible, otherwise modeling using <complex metadata> is recommended.
        - In the context where metadata such as functional annotations must be included in GFF3 column 9, the general format we would suggest and has been adopted in Apollo (https://github.com/gmod/apollo) and Artemis (https://github.com/sanger-pathogens/Artemis) for GO (go_annotations), Gene Product (gene_product), and Provenance (provenance) annotations is <type>=<type annotation>;. Each type is only included once, but can include multiple type annotations. Note that <type annotation> is URL encoded.
        - <type_annotations> are URL encoded and of the format: rank=<rankA>;<key1>=<value1A>;<key2>=<value2A>,rank=<rankB>;<key1>=<value1B>;<key2>=<value2B> . With multiple annotations, provide a rank to indicate which would go first, though this may not always be relevant. Multiple annotations are comma-delimited. Multiple key/value pairs are semicolon delimited.
        - GO terms should be updated annually - you might not want to include this as 'static' information.
- **Validation:**
    - <type> should be lower-case and be of the form <type1>=<type1 annotations>;<type2>=<type2 annotations>; etc.
    - <type annotation> entries are URL encoded
    - A <type> can have multiple <type annotations>, which are separated by a comma (url-encoded %3B).
    - Each <type annotation> has multiple key-value pairs, separated by a semi-colon (url-encoded as %2C).

- \<rank\> is not necessary, but it is preferred if more than one annotation for a type exists.
- \<type\>, \<rank\>, \<key\> should all be lower-case.
- **Example** in Supplementary data

## 17. Modeling hierarchical relationships of a protein-coding gene

- **Change level:** Recommendation only
- **Summary:** The primary purpose of GFF3 is to model gene structure.
- **Rationale:** We wish to provide a standard way to render this type of information as there are many valid ways to render the same protein-coding gene.
- **Proposed changes to specification:** None
- **Best practices:**
    - Strongly encourage only one parent per feature. However, parsers and validators should still support multiple parents per feature, in particular for elegance and backwards compatibility and to support more "non-standard" protein structures especially within non-eukaryotic organisms.
    - Edge cases that can't follow this standard recommendation exist, e.g. ribosome slippage, trans-splicing, features split across scaffolds due to assembly problems - further recommendations need to be developed.
    - Sort order. This pertains to having child features come after parent features in the gff; most loaders don't care about order of the features by coordinates. Child features should be listed after parent features.
    - Child coordinates that are not contained within parent coordinates often indicate an error and should trigger a warning in a gff3 validator.
    - The gff3 format assumes that parent and child features have a part_of relationship type.
    - There can be a ### directive between gene models.
    - Do not list multiple values in column 1 (for features split across scaffolds)
    - Polypeptide features are not required or recommended
    - Introns can be annotated, but are not necessary and are implied.
    - Type should be specified and validated as part of the sequence ontology cv terms (see also notes on column 3, type, above): http://www.sequenceontology.org/miso
- **Validation:**
    - Entries that are non-parent entries should have a valid parent entry via the ID.
    - IDs should be internally resolvable.
- **Example** in Supplementary data

18. Pragmas

Pragmas (also called directives) provide information about the entire dataset represented in the GFF3 document. Pragma lines begin with ##. Here, we suggest modifying the definition of two pragmas in the GFF3 specification.

- **Dbxref**
    - This pragma is optional.
    - Format: ##dbxref=<URI>
    - **Example**: ##dbxref=ncbiprotein:CAA71118.1; (which resolves to https://identifiers.org/ncbiprotein:CAA71118.1)
    - Addendum: This requires that the database providing the xref register at and obtain a namespace from Identifiers.org.
- **Ontology URIs**
    - In the current GFF3 specification, ontology URIs, for example ##feature-ontology URI, can be specified via cv URLs (e.g. http://song.cvs.sourceforge.net/*checkout*/song/ontology/sofa.obo?revision=1.6). These URLs should be avoided. Instead, we recommend using the official OBO version IRI PURLs, for example http://purl.obolibrary.org/obo/so.obo.
    - Example: ## feature-ontology http://purl.obolibrary.org/obo/so.obo
- **Species**
    - The current specification recommends using NCBI URLs to specify the species that annotations are derived from in the ##species pragma. We recommend using an OBO CURIE, instead.
    - Example: ##species NCBITaxon:9606

# Conclusions and Future Work

These recommendations are targeted at ameliorating challenges with protein-coding gene models represented on a common coordinate system. To complement the recommendations described by this paper, an open source software validator needs to be developed by the community that can be extended as new issues are reported by users. Future work on GFF3 specifications and recommendations should include modeling miRNAs and QTL data and addressing features located in a pan genome coordinate space. The GFF3 specification would also benefit from pragmas for describing provenance of and/or workflow used to generate the data. Given the often tight coupling of FASTA and GFF3 data, there would be a benefit in outlining best practices for FASTA definition lines to complement the GFF3 specification. We request the genomics and bioinformatics community to utilise the AgBioData GFF3 recommendation github repository (https://github.com/NAL-i5K/AgBioData_GFF3_recommendation) to actively contribute to the development of the GFF3 recommendations. This update to the original specification creates a framework for the community to collaborate and update the venerable GFF3 file format to handle feature types and data integration challenges yet unseen.

# Acknowledgements

We thank Margaret Woodhouse for the inspiration for this working group, Vamsi Kodali and Terence Murphy for input on various aspects of the GFF3 specifications, and many other contributors from AgBioData and the larger research community who have provided feedback on earlier versions.

This research was supported in part by the US. Department of Agriculture, Agricultural Research Service. Mention of trade names or commercial products in this publication is solely for the purpose of providing specific information and does not imply recommendation or endorsement by the U.S. Department of Agriculture.

The U.S. Department of Agriculture prohibits discrimination in all its programs and activities on the basis of race, color, national origin, age, disability, and where applicable, sex, marital status, familial status, parental status, religion, sexual orientation, genetic information, political beliefs, reprisal, or because all or part of an individual's income is derived from any public assistance program. (Not all prohibited bases apply to all programs.) Persons with disabilities who require alternative means for communication of program information (Braille, large print, audiotape, etc.) should contact USDA's TARGET Center at (202) 720-2600 (voice and TDD). To file a complaint of discrimination, write to USDA, Director, Office of Civil Rights, 1400 Independence Avenue, S.W., Washington, D.C. 20250-9410, or call (800) 795-3272 (voice) or (202) 720-6382 (TDD). USDA is an equal opportunity provider and employer.

# Citations

# Supplementary Data: Examples for columns, attributes, and pragmas in the GFF3 format with working group recommendations

**Seqid pragma (Column 1)**

Example of an alias table pragma for an NCBI alias table:

##alias-table

https://ftp.ncbi.nlm.nih.gov/genomes/all/GCF/000/005/005/GCF_000005005.2_B73_RefGen_v4/GCF_000005005.2_B73_RefGen_v4_assembly_report.txt

Example of a custom alias table pragma:

##alias-table Sequence-name Refseq-accession

chr1    NC_024459.2

chr2    NC_024460.2

**Source (column 2)**

gene

##gff-version 3

##Source=ID=maker_ITAG, Description="Solgenomics (SGN) tomato genome annotation pipeline based on Maker, Mikado and AHRD. Details in methods of genome publication", Software="https://doi.org/10.1101/767764"

SL4.0ch00    maker_ITAG    gene    93750    94430    .    +    .    ID=gene:Solyc00g500001.1;Alias=Solyc00g500001;Name=Solyc00g500001.1

mRNA

##gff-version 3

##Source=ID=maker_ITAG, Description="Solgenomics (SGN) tomato genome annotation pipeline based on Maker, Mikado and AHRD. Details in methods of genome publication", Software="https://doi.org/10.1101/767764"

SL4.0ch00    maker_ITAG    mRNA    93750    94430    .    +    .    ID=mRNA:Solyc00g500001.1.1;Parent=gene:Solyc00g500001.1;Name=Solyc00g500001.1.1;Note=Retrovirus-related Pol polyprotein from transposon TNT 1-94

**Source pragma**:

##Source=<ID=BestRefSeq,Description="RefSeq transcript that underwent manual inspection", Software="URL here">

**Type (column 3, example is from a RefSeq gff3 file)**

NW_023276341.1 Gnomon gene    52942    57885    .    -    .    ID=gene-LOC118063598;Dbxref=GeneID:118063598;Name=LOC118063598;gbkey=Gene;gene=LOC118063598;gene_biotype=protein_coding

NW_023276341.1 Gnomon mRNA    52942    57885    .    -    .    ID=rna-XM_035081708.1;Parent=gene-LOC118063598;Dbxref=GeneID:118063598,Genbank:XM_035081708.1;Name=XM_035081708.1;gbkey=mRNA;gene=LOC118063598;model_evidence=Supporting evidence includes similarity to: 2 Proteins%2C and 100%25 coverage of the annotated genomic feature by RNAseq alignments%2C including 5 samples with support for all annotated introns;product=uncharacterized LOC118063598;transcript_id=XM_035081708.1

NW_023276341.1 Gnomon exon    57433    57885    .    -    .    ID=exon-XM_035081708.1-1;Parent=rna-XM_035081708.1;Dbxref=GeneID:118063598,Genbank:XM_035081708.1;gbkey=mRNA;gene=LOC118063598;product=uncharacterized LOC118063598;transcript_id=XM_035081708.1

NW_023276341.1 Gnomon	exon	52942	57315	.	-	.	ID=exon-XM_035081708.1-2;Parent=rna-XM_035081708.1;Dbxref=GeneID:118063598,Genbank:XM_035081708.1;gbkey=mRNA;gene=LOC118063598;product=uncharacterized LOC118063598;transcript_id=XM_035081708.1

NW_023276341.1 Gnomon	CDS	53079	57215	.	-	0	ID=cds-XP_034937599.1;Parent=rna-XM_035081708.1;Dbxref=GeneID:118063598,Genbank:XP_034937599.1;Name=XP_034937599.1;gbkey=CDS;gene=LOC118063598;product=uncharacterized protein LOC118063598;protein_id=XP_034937599.1

**Start, end (column 4,5)**

mRNA
##gff-version 3
# organism zea mays
Chr1  gramene  mRNA  44289  49837  .  +  .  ID=420346;Name=Zm00001d027230_T001;Parent=4711

Bacterial CDS (taken from the SO specification)
##gff-version 3.1.26
# organism Enterobacteria phage f1
# Note Bacteriophage f1, complete genome.
J02448 GenBank region 1  6407 .  +  .  ID=J02448;Name=J02448;Is_circular=true;
J02448 GenBank CDS   6006 7238 .  +  0  ID=geneII;Name=II;Note=protein II;

SNP
##gff-version 3
##date Fri Dec 11 11:41:25 2020
# organism Arachis ipaensis
Araip.B01 PolymorphicArray SNP 2975884 2975884 . . . Name=AX-176823085;ID=21305;origin=A.ipaensis;alleles=A%2FG

**Score pragma (column 6)**
Format:
##Score name="[name/calculated-by]";min=[min-value]; max=[max-val];best=[lower/higher]

Example: score is the AED (Annotation Edit Distance) for a gene model feature, generated by MAKER-P.
##gff-version 3
# organism zea mays
##Score name="AED (Annotation Edit Distance) score"; min=0;max=1;best=lower
chr1 NAM mRNA 107080 108196 **0.38** - .
ID=45221;Parent=Zm00001e000004;transcript_id=Zm00001e000004_T001

**Phase (column 8).**
In this example, the phase is incorrect, but the correct sequence can be calculated by finding the longest ORF.
QKKF01020412.1  EVM	gene	1371	1574	.	+	.
    ID=evm.TU.Contig10112.1;Name=EVM prediction Contig10112.1
QKKF01020412.1  EVM	mRNA	1371	1574	.	+	.
    ID=evm.model.Contig10112.1;Parent=evm.TU.Contig10112.1;Name=EVM prediction Contig10112.1
QKKF01020412.1  EVM	exon	1371	1574	.	+	.
    ID=evm.model.Contig10112.1.exon1;Parent=evm.model.Contig10112.1

```
QKKF01020412.1   EVM      CDS       1371     1574       .         +          1
         ID=cds.evm.model.Contig10112.1;Parent=evm.model.Contig10112.1
```

Here are the CDS and protein sequence calculated using GFF3 phase 1, which is incorrect (as seen by the premature stop codons in the protein sequence):

>cds.evm.model.Contig10112.1
AGCTCGGGTGGTAATGGCATGTCGCAATTTGGAAAAAGCGGACGAGGCGGCCAAAGATATAAGGAAAACGCTGG
AAGGGGTTGAAGGTGTAGGACAAATCACTGTGAAGCATCTCGATCTGTCATCATTGTCATCTGTCAGAACCTGTGC
CGAACAACTTCTCAAAGAAGAACCAAACATACATTTATTGATTAACAATGCTG

>evm.model.Contig10112.1
SSGGNGMSQFGKSGRGGQRYKENAGRG*RCRTNHCEASRSVIIVICQNLCRTTSQRRTKHTFID*QC

Here are the correct CDS and protein sequence, calculated by finding the longest ORF.
>cds.evm.model.Contig10112.1
GAGCTCGGGTGGTAATGGCATGTCGCAATTTGGAAAAAGCGGACGAGGCGGCCAAAGATATAAGGAAAACGCTG
GAAGGGGTTGAAGGTGTAGGACAAATCACTGTGAAGCATCTCGATCTGTCATCATTGTCATCTGTCAGAACCTGTG
CCGAACAACTTCTCAAAGAAGAACCAAACATACATTTATTGATTAACAATGCTG

>evm.model.Contig10112.1
ARVVMACRNLEKADEAAKDIRKTLEGVEGVGQITVKHLDLSSLSSVRTCAEQLLKEEPNIHLLINNA

The correct gff:

```
QKKF01020412.1   EVM      gene      1371     1574       .         +          .
         ID=evm.TU.Contig10112.1;Name=EVM prediction Contig10112.1
QKKF01020412.1   EVM      mRNA      1371     1574       .         +          .
         ID=evm.model.Contig10112.1;Parent=evm.TU.Contig10112.1;Name=EVM prediction Contig10112.1
QKKF01020412.1   EVM      exon      1371     1574       .         +          .
         ID=evm.model.Contig10112.1.exon1;Parent=evm.model.Contig10112.1
QKKF01020412.1   EVM      CDS       1371     1574       .         +          0
         ID=cds.evm.model.Contig10112.1;Parent=evm.model.Contig10112.1
```

Correct CDS and protein sequence, calculated from GFF and genome fasta:
>cds.evm.model.Contig10112.1
GAGCTCGGGTGGTAATGGCATGTCGCAATTTGGAAAAAGCGGACGAGGCGGCCAAAGATATAAGGAAAACGCTG
GAAGGGGTTGAAGGTGTAGGACAAATCACTGTGAAGCATCTCGATCTGTCATCATTGTCATCTGTCAGAACCTGTG
CCGAACAACTTCTCAAAGAAGAACCAAACATACATTTATTGATTAACAATGCTG

>evm.model.Contig10112.1
ARVVMACRNLEKADEAAKDIRKTLEGVEGVGQITVKHLDLSSLSSVRTCAEQLLKEEPNIHLLINNA

**Phase pragma (column 8)**
##Translation-table 5 scaffold1,scaffold2

**Attributes (column 9): ID**

```
NW_023276341.1  Gnomon  gene    52942   57885   .   -   .   ID=gene-
LOC118063598;Dbxref=GeneID:118063598;Name=LOC118063598;gbkey=Gene;gene=LOC118063598;gene_biotype=protein
_coding
NW_023276341.1  Gnomon  mRNA    52942   57885   .   -   .   ID=rna-XM_035081708.1;Parent=gene-
LOC118063598;Dbxref=GeneID:118063598,Genbank:XM_035081708.1;Name=XM_035081708.1;gbkey=mRNA;gene=LOC1
18063598;model_evidence=Supporting evidence includes similarity to: 2 Proteins%2C and 100%25 coverage of the annotated
genomic feature by RNAseq alignments%2C including 5 samples with support for all annotated introns;product=uncharacterized
LOC118063598;transcript_id=XM_035081708.1
NW_023276341.1  Gnomon  exon    57433   57885   .   -   .   ID=exon-XM_035081708.1-1;Parent=rna-
XM_035081708.1;Dbxref=GeneID:118063598,Genbank:XM_035081708.1;gbkey=mRNA;gene=LOC118063598;product=unch
aracterized LOC118063598;transcript_id=XM_035081708.1
NW_023276341.1  Gnomon  exon    52942   57315   .   -   .   ID=exon-XM_035081708.1-2;Parent=rna-
XM_035081708.1;Dbxref=GeneID:118063598,Genbank:XM_035081708.1;gbkey=mRNA;gene=LOC118063598;product=unch
aracterized LOC118063598;transcript_id=XM_035081708.1
NW_023276341.1  Gnomon  CDS     53079   57215   .   -   0   ID=cds-XP_034937599.1;Parent=rna-
XM_035081708.1;Dbxref=GeneID:118063598,Genbank:XP_034937599.1;Name=XP_034937599.1;gbkey=CDS;gene=LOC118
063598;product=uncharacterized protein LOC118063598;protein_id=XP_034937599.1
```

### Attributes (col 9): Name

```
KZ308124.1  ladful_OGSv1.0  mRNA    2340580 2344890 .   +   .   ID=LFUL019771-RA;Name=Odorant
receptor 1;Parent=LFUL019771;product=Odorant receptor 1;transcript_id=gnl|J437|LFUL019771-RA;
```

### Attributes (col 9): Alias

In the case of pax6a (http://zfin.org/ZDB-GENE-990415-200) there are multiple "aliased names" associated with it under previous names: pax-a Pax6.1 pax6 pax[zf-a] (1) paxzfa zfpax-6a cb280 (1) etID309716.25 (1) fc20e07 wu:fc20e07 (1) zfpax-6b

```
##gff-version 3
25      ZFIN    gene    15029041        15049781        1   -   .
        ID=403398;Name=pax6a;gene_id=ZDB-GENE-990415-200;Alias=pax-a, Pax6.1, pax[zf-a], paxzfa
25      ZFIN    mRNA    15029041        15049694        .   -   .   ID=403399;Name=pax6a-
204;Parent=403398
```

### Attributes (col 9): Dbxref

- SNP example:
    - db_xref=dbSNP:rs133073; (resolves to https://www.ncbi.nlm.nih.gov/snp/rs133073?horizontal_tab=true via http://identifiers.org/dbSNP:rs133073)
- Protein-coding gene example:
    - Dbxref=MaizeGDB.locus:12098 (resolves to https://www.maizegdb.org/gene_center/gene/12098 via https://identifiers.org/maizegdb.locus:12098)

### Attributes (col 9): Derives_from

Using derives_from for polypeptide features in a gene model (if absolutely necessary, modified from the GFF3 specification: https://github.com/The-Sequence-Ontology/Specifications/blob/master/gff3.md)

```
chrX  . gene         XXXX YYYY  . + . ID=gene01;name=resA
chrX  . mRNA         XXXX YYYY  . + . ID=tran01;Parent=gene01
chrX  . exon         XXXX YYYY  . + . Parent=tran01
chrX  . CDS          XXXX YYYY  . + . ID=cds01;Parent=tran01
chrX  . polypeptide  XXXX YYYY  . + . ID=poly01;Derives_from=cds01
```

Using derives_from for polycistronic genes (copied from the GFF3 specification: https://github.com/The-Sequence-Ontology/Specifications/blob/master/gff3.md)

```
chrX . gene XXXX YYYY . + . ID=gene01;name=resA
chrX . gene XXXX YYYY . + . ID=gene02;name=resB
chrX . gene XXXX YYYY . + . ID=gene03;name=resX
chrX . gene XXXX YYYY . + . ID=gene04;name=resZ
chrX . mRNA XXXX YYYY . + . ID=tran01;Parent=gene01,gene02,gene03,gene04
chrX . exon XXXX YYYY . + . ID=exon00001;Parent=tran01
chrX . CDS  XXXX YYYY . + . Parent=tran01;Derives_from=gene01
chrX . CDS  XXXX YYYY . + . Parent=tran01;Derives_from=gene02
chrX . CDS  XXXX YYYY . + . Parent=tran01;Derives_from=gene03
chrX . CDS  XXXX YYYY . + . Parent=tran01;Derives_from=gene04
```

### Attributes (col 9): Note

...;Note=This was a gene split from pax6;Note=Version 6.3a published on Sep 15, 2020

### Attributes (col 9): Ontology_term

Ontology_term="GO:0046703" (although note that evidence code should also be included; see discussion on "complex metadata")

### Attributes (col 9): Target, Gap

Representing synteny:
glyma.Wm82.gnm2.ann1.Gm01   DAGchainer   syntenic_region 227021  926129  1243.0  +  .  **Target=glyma.Wm82.gnm2.ann1.Gm08 2159073 2752670 +**; median_Ks=0.7982

Representing sequence-level alignment:
scaffold_44   blastn   match_part   36730   38665   1758   -   .   ID=33;Parent=32;**Target=scaffold_18G19.1 1 1934 +;Gap=M884 D1 M947 D1 M103**

### Complex metadata

chr9  .  mRNA  99185587  99185866  .  -  .  owner=demo@demo.com;**provenance=rank%3D1%3Bfield%3DDESCRIPTION%3Bdb_xref%3D:%3Bevidence%3DECO:0000501%3Bnote%3D["Description of provenance of an individual field within an annotation. "]%3Bbased_on%3D[]%3Blast_updated%3D2021-01-11 16:39:32.454%3Bdate_created%3D2021-01-11 16:39:32.454;**Parent=c9637c84-1c18-4320-8c58-7277ef768fd9;**go_annotations=rank%3D1%3Baspect%3DMF%3Bterm%3DGO:0004381%3Bdb_xref%3DPMID:171711%3Bevidence%3DECO:0000315%3Bgene_product_relationship%3DRO:0002327%3Bnegate%3Dfalse%3Bnote%3D["This is a made up example."]%3Bbased_on%3D["UniProt:`123141"]%3Blast_updated%3D2021-01-11**

**16:45:45.133%3Bdate_created%3D2021-01-11 16:45:45.133%2Crank%3D2%3Baspect%3DBP%3Bterm%3DGO:0000077%3Bdb_xref%3DAspGD_REF:ASPL0000000005%3Bevidence%3DECO:0000501%3Bgene_product_relationship%3DRO:0002331%3Bnegate%3Dfalse%3Bnote%3D["This was pulled from AMIGO: http://amigo.geneontology.org/amigo/gene_product/AspGD:ASPL0000108267"%2C"This is an example GO functional annotation created within Apollo"]%3Bbased_on%3D["SGD:S000002848"%2C"UniProt:11771"]%3Blast_updated%3D2021-01-11 16:37:58.871%3Bdate_created%3D2021-01-11 16:35:41.092;gene_product=rank%3D1%3Bterm%3DAfu5g12110%3Bdb_xref%3DAspGD_REF:ASPL0000000005%09%3Bevidence%3DECO:0000501%3Balternate%3Dfalse%3Bnote%3D["Created from http://amigo.geneontology.org/amigo/gene_product/AspGD:ASPL0000108267"]%3Bbased_on%3D["SGD:S000002848\t"]%3Blast_updated%3D2021-01-11 16:38:52.342%3Bdate_created%3D2021-01-11 16:38:52.342;ID=977dcd44-5d57-4eea-9127-a784b2400f3f;**orig_id=transcript:ENST00000469816;date_last_modified=2021-01-11;Name=RN7SL794P-201-00001;date_creation=2020-07-23

## Modeling hierarchical relationships of a protein-coding gene

```
##gff-version 3
##sequence-region chr9 1 138394717
chr9	.	gene	99206226	99222116	.	-	.	owner=demo@demo.com;ID=98674a30-c148-4d05-8b17-53e3d09645ac;date_last_modified=2020-04-09;Name=ALG2-002a;date_creation=2019-11-05
chr9	.	mRNA	99206226	99222116	.	-	.	owner=demo@demo.com;Parent=98674a30-c148-4d05-8b17-53e3d09645ac;ID=7dd2784f-6357-4852-9c9b-21d8483255d8;orig_id=transcript:ENST00000476832;date_last_modified=2020-03-02;Name=ALG2-002a-00001;date_creation=2019-10-25
chr9	.	CDS	99217934	99218680	.	-	0	Parent=7dd2784f-6357-4852-9c9b-21d8483255d8;ID=b0fc5d5b-65e6-41a0-a65e-7afadfe1d8d1;Name=b0fc5d5b-65e6-41a0-a65e-7afadfe1d8d1
chr9	.	exon	99221130	99222116	.	-	.	Parent=7dd2784f-6357-4852-9c9b-21d8483255d8;ID=7aa73c7e-bb44-4e86-aeb3-921340d60e45;Name=7aa73c7e-bb44-4e86-aeb3-921340d60e45
chr9	.	exon	99206226	99218846	.	-	.	Parent=7dd2784f-6357-4852-9c9b-21d8483255d8;ID=9baa5521-2f79-4493-9435-0120aa4f96d0;Name=9baa5521-2f79-4493-9435-0120aa4f96d0
chr9	.	mRNA	99217367	99218836	.	-	.	owner=demo@demo.com;Parent=98674a30-c148-4d05-8b17-53e3d09645ac;ID=6eebcdae-fb8b-42af-a178-f70a0e8ff5da;orig_id=transcript:ENST00000476832;date_last_modified=2020-04-09;Name=ALG2-002a-00003;date_creation=2020-04-09
chr9	.	CDS	99217934	99218836	.	-	0	Parent=6eebcdae-fb8b-42af-a178-f70a0e8ff5da;ID=6eebcdae-fb8b-42af-a178-f70a0e8ff5da-CDS;Name=6eebcdae-fb8b-42af-a178-f70a0e8ff5da-CDS
chr9	.	exon	99217367	99218836	.	-	.	Parent=6eebcdae-fb8b-42af-a178-f70a0e8ff5da;ID=cbf89b7d-32de-40a0-be12-1332afe2a527;Name=cbf89b7d-32de-40a0-be12-1332afe2a527
chr9	.	mRNA	99217372	99218836	.	-	.	owner=demo@demo.com;Parent=98674a30-c148-4d05-8b17-53e3d09645ac;ID=3da41f24-d14b-4f01-8e27-c15f86912acb;orig_id=transcript:ENST00000319033;date_last_modified=2020-04-09;Name=ALG2-002a-00004;date_creation=2020-04-09
chr9	.	exon	99217372	99218836	.	-	.	Parent=3da41f24-d14b-4f01-8e27-c15f86912acb;ID=b5adb98d-82c9-4b5b-a320-1c044bced6e0;Name=b5adb98d-82c9-4b5b-a320-1c044bced6e0
chr9	.	CDS	99217934	99218836	.	-	0	Parent=3da41f24-d14b-4f01-8e27-c15f86912acb;ID=3da41f24-d14b-4f01-8e27-c15f86912acb-CDS;Name=3da41f24-d14b-4f01-8e27-c15f86912acb-CDS
chr9	.	mRNA	99217367	99221956	.	-	.	owner=demo@demo.com;Parent=98674a30-c148-4d05-8b17-53e3d09645ac;ID=7cf3b8f6-7241-4145-9822-00bbb8fd4a87;orig_id=transcript:ENST00000476832;date_last_modified=2020-04-08;Name=ALG2-002a-00002;date_creation=2020-04-08
chr9	.	CDS	99221547	99221954	.	-	0	Parent=7cf3b8f6-7241-4145-9822-00bbb8fd4a87;ID=010f5cfb-3f84-4e74-9ad0-6d54c1463502;Name=010f5cfb-3f84-4e74-9ad0-6d54c1463502
chr9	.	CDS	99217934	99218836	.	-	0	Parent=7cf3b8f6-7241-4145-9822-00bbb8fd4a87;ID=010f5cfb-3f84-4e74-9ad0-6d54c1463502;Name=010f5cfb-3f84-4e74-9ad0-6d54c1463502
chr9	.	exon	99221547	99221956	.	-	.	Parent=7cf3b8f6-7241-4145-9822-00bbb8fd4a87;ID=58e7b74f-f60a-438f-a869-e8a0fc13bd3a;Name=58e7b74f-f60a-438f-a869-e8a0fc13bd3a
chr9	.	exon	99217367	99218836	.	-	.	Parent=7cf3b8f6-7241-4145-9822-00bbb8fd4a87;ID=d39ce049-51aa-48b4-a45b-ce8422b7693b;Name=d39ce049-51aa-48b4-a45b-ce8422b7693b
###
```